\documentclass[twocolumn,aps,prl,draft,showpacs]{revtex4}
\usepackage{graphicx}
\usepackage{bm}
\begin{document}
\title{Active {\it vs} passive scalar turbulence}
\author{Antonio Celani$^1$, Massimo Cencini$^2$, Andrea Mazzino$^{3,4}$
and Massimo Vergassola$^2$}
\affiliation{$^1$ CNRS, INLN, 1361 Route des Lucioles, 06560 Valbonne, 
France.\\
$^2$CNRS, Observatoire de la C\^ote d'Azur, B.P. 4229,
06304 Nice Cedex 4, France.\\
$^3$ISAC-CNR, Str. Prov. Lecce-Monteroni Km 1.200,
I--73100 Lecce, Italy.\\
$^4$INFM--Dipartimento di Fisica, Universit\`a di Genova,
Via Dodecaneso 33, I-16146 Genova, Italy.}
\date{\today}
\begin{abstract}
Active and passive scalars transported by an incompressible
two-dimensional conductive fluid are investigated. It is shown that a
passive scalar displays a direct cascade towards the small scales
while the active magnetic potential builds up large-scale structures
in an inverse cascade process.  Correlations between scalar input and
particle trajectories are found to be responsible for those dramatic
differences as well as for the behavior of dissipative anomalies.
\end{abstract}
\pacs{47.27.-i} \maketitle The transport of scalar fields by turbulent
flows is a common physical phenomenon.  The dynamics of the advecting
velocity field often depends on the transported field, which is then
dubbed {\em active}.  That is, for instance, the case of temperature,
affecting the velocity via buoyancy forces. Conversely, situations
where the flow is not influenced by the scalar, e.g. for the
concentration of a dilute tracer, are referred to as {\em
passive}. Progress has been recently made for the passive case (see
Ref.~\cite{FGV01} and references therein).  These studies put renewed
emphasis on the known field-particle duality and long known concepts
as intermittency, dissipative anomaly, direct and inverse cascades,
usually defined in terms of field characteristics (the Eulerian
description), can now be elegantly explained in terms of the
statistical properties of particle trajectories (the Lagrangian
description).  In this Letter we address the problem of relating the
Eulerian properties to the Lagrangian ones for {\em active} scalar
transport.

The active field $a({\bm x},t)$ advected by an incompressible
flow ${\bm v}({\bm x},t)$ is governed by the transport equation
\begin{equation}
\partial_t a+ {\bm v}\cdot {\bm \nabla} a =
\kappa \Delta\, a + f_a \;.
\label{eq:1}
\end{equation}
The passive field $c({\bm x},t)$ evolves {\em in the same flow} as
\begin{equation}
\partial_t c+ {\bm v}\cdot {\bm \nabla} c =
\kappa \Delta\, c + f_c \;.
\label{eq:2}
\end{equation}
The two scalars have the same diffusion coefficient $\kappa$ while
$f_a$ and $f_c$ are independent realizations of the same random
forcing, with characteristic lengthscale $l_f$.  The difference
between active and passive scalars resides in their relationship with
the flow ${\bm v}$: the active field enters the velocity dynamics
whereas the passive one does not. Here, the evolution equation for
${\bm v}$ is
\begin{equation}
\partial_t {\bm v} + {\bm v}\cdot {\bm \nabla} {\bm v} = -{\bm \nabla} p
- \Delta a {\bm \nabla} a + \nu\Delta{\bm v} \;.
\label{eq:3}
\end{equation}
Eqs.~(\ref{eq:1}) and (\ref{eq:3}) are the two-dimensional
magnetohydrodynamics equations (see, e.g., Ref.~\cite{Bis}), with the
following glossary: $a$ is the scalar magnetic potential; the magnetic
field is ${\bm b}=(-\partial_2 a,\partial_1 a)$; the coupling $-\Delta
a {\bm \nabla} a$ is the Lorentz force $({\bm \nabla} \times {\bm
b})\times {\bm b}$; the condition ${\bm \nabla}\cdot{\bm v}=0$ is
ensured by the pressure term ${\bm \nabla} p$ and $\nu$ is the
kinematic viscosity.

\begin{figure}[b]
\includegraphics[draft=false, scale=0.68, clip=true]{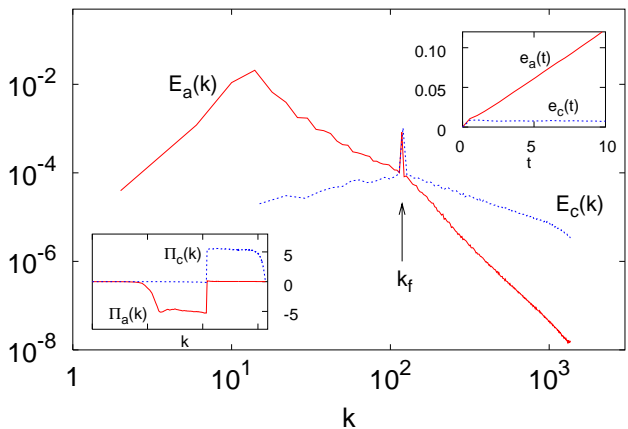} 
\caption{Power spectra of active and passive scalar variances
$E_a(k)=\pi k |\hat{a}({\bm k},t)|^2$ and $E_c(k)=\pi k |\hat{c}({\bm
k},t)|^2$.  The fluctuations of $a$ and $c$ injected at the forcing
wavelength $k_f$ flow towards smaller and larger wavenumbers,
respectively. At $k<k_f$ we observe power-law behaviors $E_{a}(k) \sim
k^{-2.0\pm 0.1}$ and $E_c(k) \sim k^{0.7\pm 0.1}$, while at $k>k_f$ we
find $E_a(k)\sim k^{-3.6 \pm 0.1}$ and $E_c(k)\sim k^{-1.4\pm 0.1}$
(for a discussion about the spectra of two- and three-dimensional
magnetohydrodynamics see, e.g.,
Refs.~\cite{Bis2,Pouq,Srid,Verm,Pouq2,Bis3}).  In the lower left
corner, the fluxes of scalar variance $\Pi_{a,c}$ out of wavenumber
$k$. Negative values indicate an inverse cascade.  In the upper right
corner, the total scalar variance $e_{a,c}(t)= \int E_{a,c}(k,t)\,dk$.
The active variance $e_a(t)$ grows linearly in time whereas $e_c(t)$
fluctuates around a finite value (see text). The rate of active to
passive scalar dissipation is $\epsilon_a/\epsilon_c \simeq 0.005$.
The data result from the numerical integration of
eqs.~(\protect\ref{eq:1}-\protect\ref{eq:3}) by a dealiased
pseudo-spectral parallel code, on a doubly periodic box of size $2\pi$
and resolution $4096^2$.  The forcing terms $f_a$ and $f_c$ are
homogeneous independent Gaussian processes with zero mean and
correlation $\langle \hat{f}_i({\bm k},t)\hat{f}_j({\bm k}',t')\rangle
=[F_0/(2\pi k_f)] \delta_{ij}\delta({\bm k}+{\bm
k}')\delta(k-k_f)\delta(t-t')$ where $i,j=a,c$.  The coefficients
$\kappa$ and $\nu$ are chosen to obtain a dissipative lengthscale of
the order of the smallest resolved scales.  All fields are set to zero
at $t=0$, and time is defined in units of eddy-turnover time
$\tau=l_f/v_{\mathrm{rms}}$ where $l_f=2\pi/k_f$.}
\label{fig:1} 
\end{figure}

Besides its physical relevance (see e.g. \cite{Bis}), this example of
active scalar is particularly interesting because of the conspicuous
differences to its passive counterpart.  Namely, whilst $a$ undergoes
an inverse cascade process, i.e. it forms structures at increasingly
larger scales \cite{Bis2}, $c$ cascades downscale (see
Fig.~\ref{fig:1}).  As a consequence, the dissipation of active scalar
fluctuations by molecular diffusivity $\epsilon_a = \kappa |{\bm
\nabla a}|^2$ vanishes in the limit $\kappa\to 0$ : no dissipative
anomaly for the field $a$.  Since the scalar variance is injected at a
constant rate $F_0$ by the source $f_a$, we have that $e_a(t)={1\over
2} \int a({\bm x},t)^2 d{\bm x}$ increases linearly in time as ${1
\over 2} F_0 t$.  On the contrary, passive scalar dissipation
$\epsilon_c = \kappa |{\bm \nabla c}|^2$ equals the input ${1 \over 2}
F_0$ and holds $c$ in a statistically stationary state (see
Fig.~\ref{fig:1}).

\begin{figure}[b]
\includegraphics[draft=false, scale=0.65, clip=true]{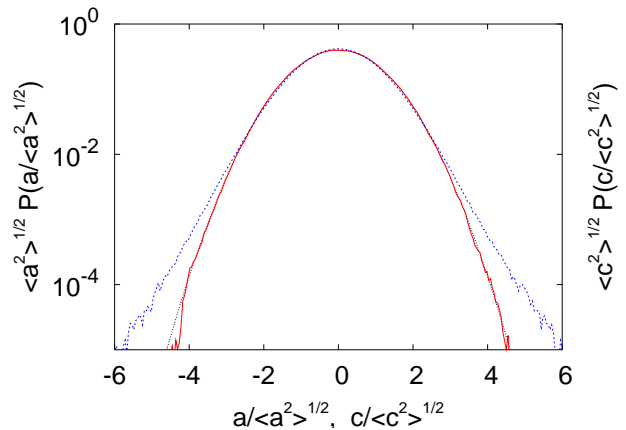} 
\caption{Pdf's of active (solid) and passive (dashed) scalar fields
normalized by their standard deviation.  The active scalar pdf is
indistinguishable from a Gaussian (dotted).}
\label{fig:2} 
\end{figure}

The probability distribution function (pdf) of $a$ is Gaussian, with
zero mean and variance $F_0 t$ (see Fig.~\ref{fig:2}).  This is a
straightforward consequence of the vanishing of active scalar
dissipation. Indeed, by averaging over the forcing statistics one
obtains: $\partial_t \langle a^{2n} \rangle = n(2n-1)F_0 \langle
a^{2n-2} \rangle$ (odd moments vanish by symmetry) whose solutions are
the Gaussian moments: $\langle a^{2n} \rangle = (2n-1)!!(F_0t)^n$.
This may be contrasted with the single-point pdf of $c$ which is
stationary and supergaussian (see Fig.~\ref{fig:2}), as it generically
happens for a passive field sustained by a Gaussian forcing in a rough
flow (see, e.g., Ref.~\cite{FGV01}).

We now focus on the main purpose of this Letter, that is to
investigate the statistical properties of active and passive scalar in
the Lagrangian setting.  The fundamental property which we will
exploit is that the equations (\ref{eq:1}) and (\ref{eq:2}) can be
formally solved in terms of particle propagators. Let ${\bm X}(s)$ be
the trajectory of a fluid particle transported by the flow ${\bm v}$
and subject to a molecular diffusivity $\kappa$, landing at point
${\bm x}$ at time $t$.  The particle moves according to the stochastic
differential equation $d{\bm X}(s) = {\bm v}({\bm
X}(s),s)ds+\sqrt{2\kappa}\,d{\bm \beta}(s)$, where ${\bm\beta}(s)$ is
a two-dimensional brownian motion. The probability density $p({\bm
y},s|{\bm x},t)$ of finding a particle at point ${\bm y}$ and time
$s \le t$ obeys the Kolmogorov equations (see, e.g., \cite{Risk}):
\begin{equation}
- \partial_s p({\bm y},s|{\bm x},t) -{\bm \nabla}_{\!y} \cdot 
[{\bm v}({\bm y},s) p({\bm y},s|{\bm x},t) ] \!=\! 
\kappa \Delta_y p({\bm y},s|{\bm x},t) \,,
\label{eq:6}
\end{equation} 
\begin{equation}
\partial_t p({\bm y},s|{\bm x},t) +{\bm \nabla}_{\!x} \cdot 
[{\bm v}({\bm x},t) p({\bm y},s|{\bm x},t) ] \!=\! 
\kappa \Delta_x p({\bm y},s|{\bm x},t) \,.
\label{eq:6bis}
\end{equation} 
The unusual minus signs in the l.h.s. of (\ref{eq:6}) are due
to the fact that particles move {\em backward} in time.  
The solution of (\ref{eq:1}) and (\ref{eq:2}) can be written in 
terms of the propagator:
\begin{eqnarray}
a({\bm x},t)&=&\int_0^t ds\int d{\bm y}\,f_a({\bm y},s) 
p({\bm y},s|{\bm x},t)\;,
\label{eq:4}\\
c({\bm x},t)&=&\int_0^t ds\int d{\bm y}\,f_c({\bm y},s) 
p({\bm y},s|{\bm x},t)\;,
\label{eq:5}
\end{eqnarray} 
as it can be directly checked by inserting (\ref{eq:4}) and
(\ref{eq:5}) in (\ref{eq:1}) and (\ref{eq:2}), respectively, and
utilizing (\ref{eq:6bis}).  It is important to notice that both the
active and the passive scalar are expressed in terms of {\em the same
propagator} $p$, since $a$ and $c$ are advected by the same velocity
field.  Eq.~(\ref{eq:4}) can be written more compactly as $a({\bm
x},t)=\langle \int_0^t f_a({\bm X}(s),s) ds \rangle_{X}$, where
$\langle \ldots\rangle_X$ denotes the average over particle
trajectories, and similarly for eq.~(\ref{eq:5}).

\begin{figure}[b]
\includegraphics[draft=false, scale=0.6, clip=true]{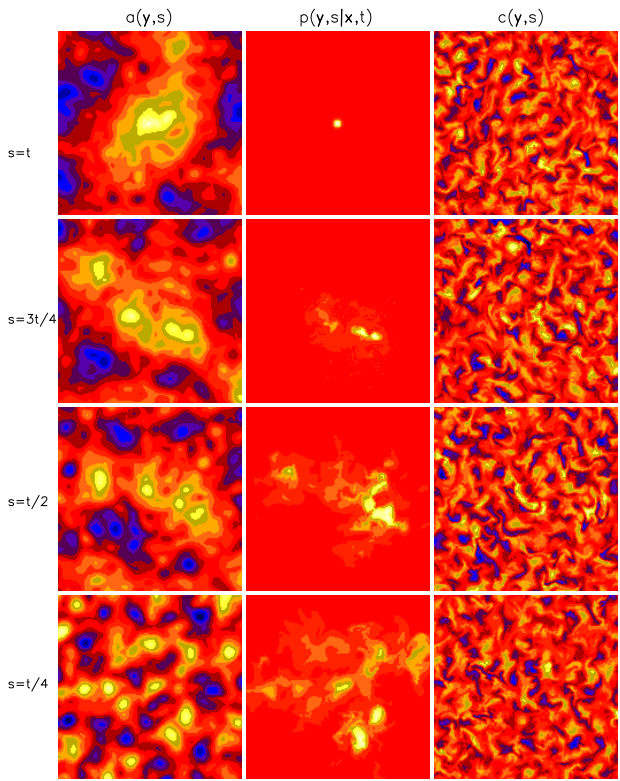} 
\caption{Time runs from bottom to top. First column: time evolution of
the active scalar field resulting from the numerical integration of
eqs.~(\protect\ref{eq:1}) and (\protect\ref{eq:3}).  Second column:
backward evolution of the particle propagator according to
eq.~(\protect\ref{eq:6}).  Third column: time evolution of the passive
scalar field in the same flow.}
\label{fig:3} 
\end{figure}
The interpretation of eqs.~(\ref{eq:4}) and (\ref{eq:5}) is that the
value of the scalar field at $({\bm x},t)$ is given by the
superposition of the input along all trajectories eventually convening
at that point and time. It is worth remarking that, even in the limit
of vanishing diffusivity, the particle propagator does not collapse
onto a single trajectory, as expected for incompressible rough flows
\cite{FGV01}.  The differences between active and passive fields are
due to the correlations between the propagator $p$ and the forcing
$f_a$, which are absent, by definition of passive transport, between
$p$ and $f_c$. Indeed, via the coupling term in eq.~(\ref{eq:3}),
$f_a$ affects the velocity field and consequently the evolution of the
propagator $p$.  To further clarify the relationship between
Lagrangian trajectories and active field forcing, it is necessary to
investigate the evolution of the particle propagator. The problem is
that $p$ evolves {\em backward} in time, according to (\ref{eq:6}) and
the condition $p({\bm y},t|{\bm x},t)=\delta({\bm y}-{\bm x})$ is set
at the final time $t$. Contrary to an usual forward integration, where
${\bm v}$ and $p$ can be advanced in parallel, a brute force backward
integration would require the storage of velocity configurations for
the whole lapse of integration which is quite unfeasible.  To overcome
this problem we devised a fast and low-memory demanding algorithm to
integrate numerically eqs.~(\ref{eq:1}), (\ref{eq:3}) and (\ref{eq:6})
\cite{alain}.

Let us sketch the basic idea of the algorithm.  We proceed inductively
and first consider the case $t=t_1=\Delta t$, i.e. a single step. The
fields $a(0)$ and ${\bm v}(0)$ and the final condition $p(t_1)$ are
given as input (we omit spatial dependences). The fields $a(t_1)$ and
${\bm v}(t_1)$ are obtained by a forward integration whose numerical
cost is denoted $c_f$.  The propagator is then evolved backward, at a
cost denoted $c_b$, to obtain $p(0)$. The total cost of the procedure
is $C_1=c_f+c_b$.  Consider now the case $t=t_N=(2^N-1)\Delta t$. Our
aim is to show that the algorithm for $2^N-1$ steps can be split into
two successive applications with $2^{N-1}-1$ steps and thus lends to a
recursive procedure.  Integrating forward $a(0)$, ${\bm v}(0)$ for
$2^{N-1}$ steps give $a(t_{N-1}+\Delta t)$ and ${\bm v}(t_{N-1}+\Delta
t)$ at a cost $2^{N-1}c_f$.  Those fields, together with $p(t_N)$, are
the input for the algorithm in the interval $[t_{N-1}+\Delta t,t_N]$,
which gives $p(t_{N-1}+\Delta t)$ at a cost $C_{N-1}$. A backward step
is finally needed to obtain $p(t_{N-1})$.  With this field and $a(0)$,
${\bm v}(0)$ we can now apply the algorithm in the first half of the
interval $[0,t_{N-1}]$ to get $p(0)$. The computational cost is
$c_b+C_{N-1}$.  Summing up all the costs we have the recursive
relation $C_N=2^{N-1}c_f+ 2C_{N-1}+c_b$, with solution
$C_N=\frac{N}{2}2^N c_f+(2^N-1)c_b$.  Comparing $C_N$ with the cost of
a forward integration of eqs.~(\ref{eq:1}),(\ref{eq:3}) and
(\ref{eq:6bis}), that is $(c_f+c_b)(2^N-1)$, we have an increase by a
factor $\approx \frac{c_f}{2(c_f+c_b)}N$, i.e. logarithmic in the
number of timesteps.  As for memory requirements, only the $N$ couples
of fields $a(t_N-t_k),{\bm v}(t_N-t_k), k=1,\ldots,N$ need to be
stored -- again a logarithmic factor.

\begin{figure}[b]
\includegraphics[draft=false, scale=0.68, clip=true]{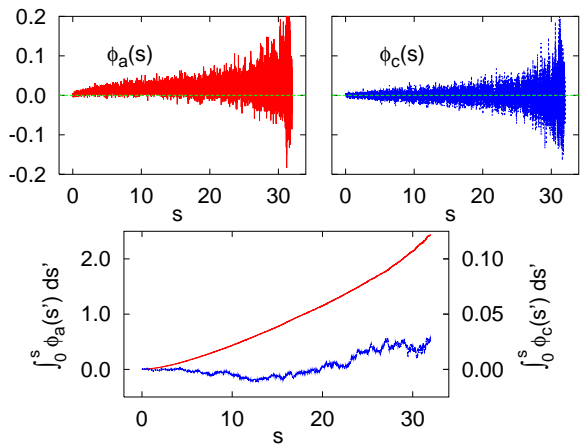} 
\caption{Top: $\phi_{a,c}(s)=\int d{\bm y}\,f_{a,c}({\bm y},s) p({\bm
y},s|{\bm x},t)$. The two graphs have the same scale on the vertical
axis. Here, $t=32$. Bottom: time integrals $\int_0^s \phi_{a}(s')\,ds'$
(upper curve) and $\int_0^s \phi_{c}(s')\,ds'$ (lower curve). Note
the different scale on the vertical axis. Recall that $\int_0^t
\phi_{a}(s')\,ds' =a({\bm x},t)$ (and similarly for $c$).  }
\label{fig:4} 
\end{figure}

A typical evolution of the propagator is shown in the central column
of Fig.~\ref{fig:3}.  We can now reconstruct the time sequence of the
forcing contributions $\phi_{a,c}(s)=\int d{\bm y}\,f_{a,c}({\bm y},s)
p({\bm y},s|{\bm x},t)$ which, integrated over $s$, gives the
amplitude of the scalar fields according to (\ref{eq:4}) and
(\ref{eq:5}).  As shown in Fig.~\ref{fig:4}, the time series of
$\phi_a(s)$ and $\phi_c(s)$ are markedly different.
In the active scalar case, the sequence is strongly skewed towards
positive values at all times. This signals that the trajectories
preferentially select regions where $f_a$ has a positive sign, summing
up forcing contributions to generate a typical variance of $a$ of the
order $F_0t$.  Conversely, $f_c$ can be positive or negative with
equal probability on distant trajectories, and the time integral in
eq.~(\ref{eq:5}) averages out to zero for $|s-t|>\tau$
($\tau=l_f/v_{\mathrm{rms}}$ is the eddy-turnover time). Therefore,
typically $c^2 \sim F_0 \tau$.  The cumulative effect of the
correlation between forcing and propagator is even more evident from
the relation $\int_0^s \phi_a(s')\,ds'=\int d{\bm y}\, a({\bm
y},s)p({\bm y},s|{\bm x},t)$ derived from (\ref{eq:1}) and
(\ref{eq:6}). As shown in Fig.~\ref{fig:4}, the growth of the scalar
variance is thus related to a strong spatial correlation between the
propagator and the active field.  That is further evidenced by
comparing the first and the second column of Fig.~\ref{fig:3}: the
distribution of particles closely follows the distribution of
like-sign active scalar. This amounts to say that large-scale scalar
structures are built out of smaller ones which ``coalesce'' together
\cite{Bis2}.  This situation has to be contrasted with the absence of
large-scale correlations between the propagator and the passive scalar
field (second and third column of Fig.~\ref{fig:3}).

As remarked previously, the absence of dissipative anomaly is closely
related to the onset of an inverse cascade. Let us now discuss this
issue from a Lagrangian viewpoint and consider the squared active
field $a^2$. On one hand, it can be written as the square of
(\ref{eq:4}). On the other hand, one can multiply (\ref{eq:1}) by $2
a$ to obtain the equation $\partial_t a^2 +{\bm v}\cdot\nabla a^2 =
\kappa \Delta a^2 +2af_a -2\epsilon_a$ and solve it for $\epsilon_a=0$ in
terms of particle trajectories. The comparison of the two previous
expressions yields
\begin{eqnarray}
\int_0^t\hspace{-3pt} ds \hspace{-2pt}\int_0^t \hspace{-3pt}ds'
\hspace{-3pt}\int\hspace{-8pt}\int\hspace{-3pt}
f_a({\bm y},s) f_a({\bm y}',s')
p({\bm y},s|{\bm x},t)p({\bm y}',s'|{\bm x},t)&=& \nonumber\\
\int_0^t\hspace{-3pt} ds \hspace{-2pt}\int_0^t \hspace{-3pt}ds'
\hspace{-3pt}\int\hspace{-8pt}\int\hspace{-3pt}
f_a({\bm y},s) f_a({\bm y}',s')
p({\bm y}',s';{\bm y},s|{\bm x},t)\,,& &
\label{eq:7}
\end{eqnarray}
where $p({\bm y}',s';{\bm y},s|{\bm x},t)= p({\bm y},s|{\bm
x},t)p({\bm y}',s'|{\bm y},s)$ denotes the probability that a
trajectory ending in $({\bm x},t)$ were in $({\bm y},s)$ and $({\bm
y}',s')$.  Integration over ${\bm y}$ and ${\bm y}'$ is implied.  
To obtain the r.h.s. of (\ref{eq:7}) consider the expression  
$a^2({\bm x},t)=2 \int_0^t ds \int d{\bm y}\, f_a({\bm y},s)
a({\bm y},s)$, insert 
$a({\bm y},s)=\int_0^s ds' \int d{\bm y}'\, f_a({\bm y}',s')$, i.e.
eq.~(\ref{eq:4}) evaluated at time $s$, and exploit the symmetry under 
exchange of $s$ and $s'$ to switch from a time-ordered form to
a time-symmetric one to get rid of the factor two.

Notice as a side remark that by integrating the r.h.s. of (\ref{eq:7})
over ${\bm x}$, exploiting the normalization $\int d{\bm x} p({\bm
y},s|{\bm x},t)=1$ (as follows from(\ref{eq:6bis})), and averaging
over the forcing statistics, one obtains $\langle a^2
\rangle=F(0)t$. The result can be generalized to show that $\langle
a^{2n} \rangle=(2n-1)!! (F(0)t)^n$, in agreement with the Eulerian
argument for the Gaussianity of the single-point pdf of $a$ given
above.

Eq.~(\ref{eq:7}) can be recast as $\langle \int_0^t f_a({\bm
X}(s),s)ds \rangle_X^2= \langle [\int_0^t f_a({\bm X}(s),s) ds]^2
\rangle_X$, leading to the conclusion that $\int_0^t f_a({\bm
X}(s),s)ds$ is a {\em non-random} variable over the ensemble of
trajectories.  This result has a simple interpretation: the absence of
dissipative anomaly (i.e. $\epsilon_a=0$) is equivalent to the
property that along any of the infinite trajectories ${\bm X}(s)$
ending in $({\bm x},t)$ the quantity $\int_0^t f_a({\bm X}(s),s)ds$ is
exactly the same, and equals $a({\bm x},t)$. Therefore, a single
trajectory suffices to obtain the value of $a({\bm x},t)$ contrary to
the passive case, where different trajectories contribute disparate
values of $\int_0^t f_c({\bm X}(s),s)ds$, with a typical spread
$\epsilon_c t$. In this case, only the average over all trajectories
yields the correct value of $c({\bm x},t)$.

The previous property extends to multiple points. Proceeding as for
(\ref{eq:7}), one can show that $m$ trajectories are enough to
calculate the product of arbitrary powers of $a$ at $m$ different
points. In particular, structure functions $\langle\left(a({\bm r})-
a({\bm 0})\right)^n\rangle$ for any order $n$ can be shown to involve
only two trajectories. This should be contrasted with the passive
scalar, where the number of trajectories increases with $n$ and this
is at the core of anomalous scaling of the field \cite{FGV01}. Indeed,
we do not observe any anomalous scaling for the structure functions of
the active field $a$ and the pdf's of its increments are self-similar.
Note that the same phenomenon and a lack of dissipative anomaly were
also found for a passive scalar in a compressible flow \cite{GV}.  In
that case, the ensemble of trajectories collapses onto a unique path,
fulfilling in the simplest way the constraint~(\ref{eq:7}) and its
multi-point analogs. Here, even though the trajectories do not
collapse, the constraints are satisfied due to the subtle correlation
between forcing and trajectories peculiar to the active case.

We conclude with the following open issue.  It is known that passive
correlation functions are controlled by the asymptotically dominant
terms in particle propagators (see, e.g., Ref.~\cite{FGV01}). Is this
also true for active scalars~?  The correlations between propagator
and input in (\ref{eq:4}) suggest this be not the general rule. For
some specific systems, those correlations might however be such that
the asymptotics of the propagator still stands out at large times
\cite{Itamar,Matsu}.  Clarifying the dynamical conditions controlling
this phenomenon will be the subject of future work.

We are grateful to A.~Noullez and I.~Procaccia for useful discussions.
This work has been supported by the EU under the contract
HPRN-CT-2000-00162, and by Indo-French Center for Promotion of
Advanced Research (IFCPAR~2404-2).  AM has been partially supported by
Cofin2001 (prot.2001023842).  Numerical simulations have been
performed at IDRIS (project 021226) and at CINECA (INFM parallel
computing initiative).

\end{document}